\documentclass[a4paper,fleqn]{sc}

\usepackage{amsmath,amsfonts} 
\usepackage[svgnames]{xcolor} 
\usepackage{graphicx}
\usepackage{caption}
\usepackage[authoryear]{natbib}

\def\tsc#1{\csdef{#1}{\textsc{\lowercase{#1}}\xspace}}
\tsc{WGM}
\tsc{QE}
\tsc{EP}
\tsc{PMS}
\tsc{BEC}
\tsc{DE}

\usepackage{arydshln}
\usepackage{multirow}
\usepackage{pdflscape}
\usepackage{tikz}
\usepackage{comment}
\usepackage{nicefrac}

\usetikzlibrary{positioning}
\usetikzlibrary{patterns,arrows,decorations.pathreplacing}
\usetikzlibrary{backgrounds}

\usepackage[authoryear]{natbib}

\usepackage{ragged2e}

\usepackage{subfigure}
\usepackage{booktabs}
\definecolor{amundi_blue}{RGB}{0,176,240}
\definecolor{amundi_dark_blue}{RGB}{0,28,75}
\definecolor{darkblue}{rgb}{0.0, 0.0, 0.55}
\usepackage{hyperref}
\hypersetup{
    colorlinks=true,
    citecolor=darkblue,
    linkcolor=darkblue,
    filecolor=darkblue,      
    urlcolor=darkblue
}

\usepackage{pdfpages}
\usepackage{bookmark}

\usepackage{threeparttable}

\begin{document}
\let\WriteBookmarks\relax
\def\floatpagepagefraction{1}
\def\textpagefraction{.001}
\shorttitle{Investor base and idiosyncratic volatility of cryptocurrencies}
\shortauthors{A. Izadyar. S. Zamani}

\title[mode = title]{Investor base and idiosyncratic volatility of cryptocurrencies}                      
\tnotemark[1,2]

\author[1]{\textcolor{black}{Amin Izadyar}}
\fnmark[1]

\author[1]{\textcolor{black}{Shiva Zamani}}
\cormark[1]
\ead{zamani@sharif.edu}
\ead[url]{http://gsme.sharif.ir/~zamani}


\address[1]{GSME, Sharif University of Technology, Teymoori Sq, Tehran, 1459973941, Iran}

\cortext[cor1]{Corresponding author: \\
\hspace*{6mm} Shiva Zamani\\
\hspace*{6mm} Associate Professor, Sharif University of Technology
}
\fntext[fn1]{
MBA Student, Sharif University of Technology, \\
\hspace*{6mm} email: amin.izadyar@gsme.sharif.edu
}

\begin{abstract}
This paper investigates how changes in investor base is related to idiosyncratic volatility in cryptocurrency markets.  For each cryptocurrency, we set change in its subreddit followers  as a proxy for the change in its investor base, and find out that the latter can significantly increase cryptocurrencies’ idiosyncratic volatility. This finding is not subsumed by effects of size, momentum, liquidity and volume and is robust to various measures of idiosyncratic volatility.
\end{abstract}

\begin{keywords}
Idiosyncratic volatility \sep Cryptocurrency  \sep Investor base  \sep Social media \sep Investor Attention
\end{keywords}

\maketitle

\section{Introduction}\label{sec:introduction}

Recent incidents during which Wallstreetbets subreddit played a major role in the short squeeze of AMC and GameStop stocks, displays the huge potential power that an increasing number of retail investors can wield in financial markets when coordinated by social media platforms. In the case of GameStop attack, in January 2021, Reddit users on the Wallstreetbets subreddit initiated a short squeeze that pushed up the stock price nearly 30 times\footnote{https://en.wikipedia.org/wiki/GameStop\char`_short\char`_squeeze}. Events like these have directed a lot of attention toward how social media platforms especially Reddit can influence the dynamics of financial markets. (\cite{corbet2015wereddit} , \cite{long2022ijustlike}, \cite{paderson2021gameon}) \newline
This phenomenon, the growing influence of social media, is perhaps more pronounced in the cryptocurrency markets. Although in his seminal paper (\cite{nakamoto2009Bitcoin}), creator of Bitcoin highlights the need for a decentralized electronic payment system that enables people to make transactions without the involvement of third parties and is immune to interference by governments or big financial institutions, but is the rising interest in cryptocurrencies a realization of the belief in this decentralized financial system or are they merely a speculative vehicle driven by influencer’s hype and social media sentiment? In a recent study \cite{auer2021distrust} utilize US survey data to identify whether retail investors are drawn to cryptocurrencies as an alternative to the conventional financial system. They don't come upon any evidence to support the idea that cryptocurrencies are sought as alternatives to the established financial system and investors in cryptocurrencies display no more concern about the reliability of cash or current banking system than the rest of the population. Moreover, \cite{lammer2020whoare} investigate the data on cryptocurrency-related structured retail products from a random sample of customers of a large European bank and discover that cryptocurrency investors are more prone to get involved with penny-stocks and stocks associated with pump and dump schemes. In another study, by examining Bitcoin’s financial characteristics and comparing them to a wide variety of assets and inspecting how Bitcoins are used, \cite{BAUR2018177} study whether Bitcoin is primarily used as an alternative currency or by contrast, as a speculative investment. They conclude that Bitcoin is largely employed as a speculative investment despite or due to its high volatility and large returns. Furthermore, other studies using a variety of econometric methods confirm the speculative bubble pattern in the cryptocurrency markets (\cite{GROBYS2021101289}, \cite{geuder2019bubble} , \cite{chaim2019bubble}, \cite{LIN2021101552}, \cite{vogiazes2019hell}). Therefore, in the speculative lottery-like environment of cryptocurrency markets where most instruments have no inherent value, social media influencers could have a massive impact on how the market behaves. An excellent illustration of this notion is Elon Musk’s tweets, as it was shown by numerous studies that they indeed move cryptocurrency markets greatly (\cite{fsecon1028730}, \cite{cary2021doge}, \cite{shahzad2022elon}, \cite{ante2021elon}).

These findings and lack of fundamental characteristics unlike stock markets, have inspired many scholars to conduct research on the role investor attention plays in cryptocurrency markets. Most of extant research uses tweeter data or the intensity of Google searches as a proxy for investor attention. Twitter is a great indicator of investor attention since it measures how actively investors talk about cryptocurrencies. \cite{ALGUINDY2021556} gathers a dataset comprised of number of tweets, retweets and favorites on 23 largest cryptocurrencies and finds that a greater level of investor attention causes larger volatility. Additionally, by considering the days on which many firms report their quarterly earnings as a proxy for attention-grabbing events, he reveals that the more investors are distracted, the less the market is volatile. 
The findings of his paper are consistent with the theory of investor attention and market volatility developed by \cite{anderi2015}. Likewise, other studies such as \cite{shen2018twitter}, \cite{suardi2020tweet} have utilized twitter data to capture investor attention and examine its relation with different cryptocurrency features. Since Bitcoin is an internet-based currency and investors often obtain information via search engines, the intensity of Google searches provides another good indicator of investor attention. In this regard, \cite{zhange2020attention} analyze the relationship of Google Trends data with both return and volatility of major cryptocurrencies and find that investor attention has significant forecasting power. Moreover, along this line of research, \cite{zhu2021attention}, \cite{DASTGIR2019160}, \cite{hang2021causality} try to establish causal relationship between investor attention proxied by Google search intensity and Bitcoin returns. In another study, \cite{IBIKUNLE2020101459} look into the connection between noise in the price discovery process of Bitcoin and investor attention measured by Google trends data and discover that higher investor attention is associated with more uninformed trading activity.

Attention is a scarce cognitive resource (\cite{kahneman1973attention}), and when faced with many incidents calling for their attention, investors become afflicted to attention deficiency. Hence, the resource demanding nature of selecting what securities to trade, leads to investors to be more inclined to trade the assets that catch their attention the most. In line with this, \cite{merton1987model} contends that information about securities is costly to obtain, and it is not possible for investors to follow every security in the market. As a result, investors only follow a subset of the securities available in the market and create their portfolios from these familiar instruments. This is in direct contrast with standard asset pricing models which assume infinite information processing capacity for market participants. But, when there is incomplete information and as a result, market participants are unable to hold fully diversified portfolios, investors must be compensated for securities’ idiosyncratic volatility. Other theories also suggest a positive relationship between idiosyncratic risk and expected returns (\cite{levy1978imperfect}, \cite{malkielxu2004}, \cite{campbell2001idio}). But the empirical findings have been somewhat puzzling. While some studies report a positive relationship between idiosyncratic volatility and returns (\cite{FU200924}), others report a negative relationship (\cite{ang2006vol}, \cite{ANG20091}, \cite{COTTER2015184}), and some even report an insignificant relationship (\cite{bali2008volidio}, \cite{boyer2009skew}). This anomaly has caused a great amount of debate among finance scholars. In another study, \cite{chichenea2015base} construct four proxies for investor base measuring the visibility of different stocks and find strong evidence indicating that the pricing of idiosyncratic volatility is conditional on a stock’s investor base. Despite the numerous research conducted on stock markets, little attention has been paid to the idiosyncratic volatility of cryptocurrencies. In one study, \cite{ZHANG2020101252} investigate the pricing of idiosyncratic risk and find that it is indeed a significant pricing factor in the cryptocurrency markets and is positively related to expected returns. In another paper, \cite{yao2021attention}, using Google Trends data as a proxy for investor attention, reveal that higher levels of investor attention cause lower levels of idiosyncratic volatility and they are linked together through the channel of liquidity.

This paper makes use of different cryptocurrencies’ time series of subreddit followers. changes in number of followers can be regarded as a proxy for changes in investor base of each cryptocurrency or a measure of investor attention. By forming a vast dataset of cryptocurrency prices and other characteristics, we find that changes in subreddit followers are positively related with idiosyncratic volatility even after controlling for various variables. Findings of this paper contribute to the growing literature of idiosyncratic risk in cryptocurrency markets and the role of social media and investor attention and .

\section{Methodology}\label{sec:methodology}

Trading data for all cryptocurrencies were collected from Coinmarketcap.com which is probably the most reliable source of data for studies focusing on cryptocurrency market (
\cite{LIU2020299}, \cite{liutsv2022riskfac}). Coinmarketcap tracks cryptocurrency prices on more than 200 major exchanges and makes available open, close, high, and low prices in addition to market capitalization and volume data on a daily frequency. Prices are calculated as the volume weighted averages of all quoted prices in different exchanges. To be listed on Coinmarketcap, a cryptocurrency needs to conform to a set of criteria. For instance, it needs to be traded on a public exchange with an application programming interface (API) quoting the last trade price, the last 24-hour trading volume. The nonzero trading volume on at least one supported exchange is necessary to find the price for the cryptocurrency. In order to avoid survivorship bias, we used the data of both active and defunct cryptocurrencies. Our dataset covers the period from January 2014 to June 2022. The trading volume data became available in the last week of 2013, and thus our sample period starts from the beginning of 2014. Following the procedure of \cite{liutsv2022riskfac}, we require that the coins have information on price, volume, and market capitalization, and exclude stablecoins and coins with market capitalization of less than \$ 1,000,000. All the cryptocurrencies that pass the described filters, are used in the construction of asset pricing factors. 

Following the previous literature, we assume that a cryptocurrency return is driven by a common factor and its own specific shocks (\cite{ZHANG2020101252}). Since idiosyncratic volatility is model dependent, we use two different asset pricing models, namely CAPM and Fama-French three-factor model. 
The following formula represents the traditional CAPM adjusted for cryptocurrency markets, which we call cryptocurrency CAPM, thereafter.
\begin{equation}
 R_{i,d} - r_{f,d} = \alpha_{i,d} + \beta_{i,d}\:\textit{MRKT}_d + \epsilon_{i,d}. 
 \label{CAPM}
\end{equation}
Here $d$ refers to day $d$ of a given month, $r_{f,d}$ is the riskfree rate in , $R_{i,d}$, $\alpha_{i,d}$, and $\beta_{i,d}$ are the return, alpha, and beta of cryptocurrency $i$, respectively. $\textit{MRKT}_d$ is the cryptocurrency excess market return in day $d$, and  $\epsilon_{i,d}$ is the residual term.
We construct the cryptocurrency market return (MKRT) as the value-weighted return of all underlying available coins, as (\cite{liu2020riskreturn} suggests. MKRT is calculated as the difference between the cryptocurrency market return and the risk-free rate derived from the one-month Treasury rate. The within-month daily returns are used to estimate equation (\ref{CAPM}). \\
The Fama-French three-factor model suggests the following alternative formula to describe the excess return of assets:
\begin{equation}
 R_{i,d} - r_{f,d} = \alpha_{i,d} + \beta_{i,d}\:\textit{MRKT}_d + s_{i,d}\: \textit{SMB}_d + w_{i,d}\: \textit{WML}_d + \epsilon_{i,d}, 
 \label{return}
\end{equation}
where the repeated notations have the same meaning as equation (\ref{CAPM}),  SMB stands for small minus big, and WML for winner minus loser. We construct the cryptocurrency SMB and WML factors following the method in \cite{FAMA19933}. Specifically, for size, each week we split the coins into three size groups by market capitalization: bottom 30\% (small), middle 40\% (middle), and top 30\% (big). We then form value weighted portfolios for each of the three groups. The size factor (SMB) is the difference between the returns of small and big portfolios.
We construct the momentum factor (WML) using three-week momentum and form the momentum factor portfolio based on the intersection of 2×3 portfolios. In particular, for each week, we first sort coins into two portfolios based on coin size, then form three momentum portfolios within each size portfolio based on past three-week returns. The first, second, and third momentum portfolios are the bottom 30\%, middle 40\%, and top 30\% of the coins based on past three-week returns. The momentum factor is constructed as:
\begin{equation}
\textit{WML} = \frac{1}{2} (Small\: high + Big\: high) -\frac{1}{2} (Small\: low + Big\: low).
\label{momentum}
\end{equation}
In table (\ref{table:correlation}) which presents the correlation of the three constructed factors, we observe that the factors are largely uncorrelated.



\begin{table}
    \centering

    \caption{Correlation matrix of risk factors}
\label{tab:09RM_Fil_GP-PwG}
\begin{tabular*}{\textwidth}{@{\extracolsep{\fill}\quad}@{}lcccc@{}}

\toprule
   & \textbf{WML} & \textbf{SMB} & \textbf{MRKT}\\ \midrule
\textbf{WML}              \textbf{}& 1                &                &              \\
\textbf{SMB}      & 0.114                & 1               &            \\
\textbf{MRKT} & -0.009                & -0.059               & 1             \\
 \bottomrule
\end{tabular*}


\begin{tablenotes}[flushleft]
\item Note: Pearson correlations between the risk factors of the three-factor model specified in equation (\ref{return}). 
\end{tablenotes}
\label{table:correlation}
\end{table}

\begin{figure}[htb]
\center{\includegraphics[width=\textwidth]{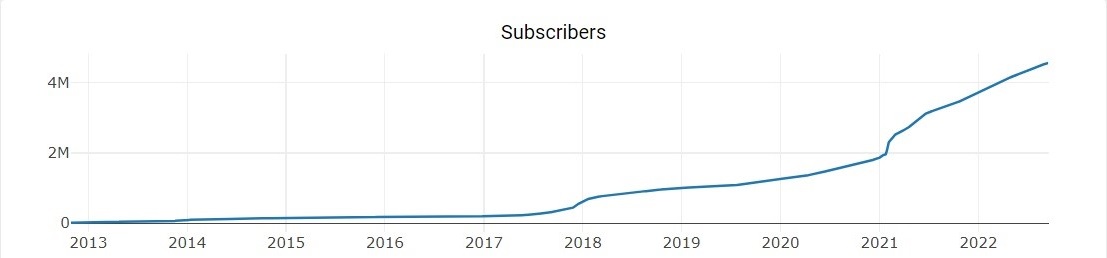}}
\caption{ Followers of Bitcoin's subreddit from 2013 to 2022. This chart is exctracted from the website subredditstats.com}
\label{fig:bitcoin followers}
\end{figure}

After constructing the required factors, we run both asset pricing models (equations (\ref{CAPM}) and (\ref{return})) and calculate the residual terms. We then measure the monthly idiosyncratic volatility for cryptocurrency $i$ as the standard deviation of the residuals:
\begin{equation}
\textit{IVOL}_{i,t} = \sqrt{Var(\epsilon_{i,d,t})}.
 \label{ivol}
\end{equation}
One of the most important variables used in this study is the time series of subreddit followers. Reddit is a social media forum where content is socially created and promoted by site members through voting and has become a popular platform for blockchain enthusiasts to interact with one another. The site is composed of hundreds of subcommunities, known as subreddits and each subreddit has a dedicated topic, such as technology, politics, a movie or a video game. Most cryptocurrencies have an active community in Reddit. Members of each community tend to exchange their ideas on the latest incidents regarding that cryptocurrency on the subreddit dedicated to it. Thus, the number of followers of a cryptocurrency’s subreddit could be a good proxy to measure its investor base. \\
By using Coinmarketcap’s API, we managed to fetch every cryptocurrency’s subreddit address. The website subredditstats.com provides many useful statistics for almost any active subreddit. In Figure (\ref{fig:bitcoin followers}) one can see the growth of Bitcoin's subreddit followers through time.
Although it creates some look ahead bias in our analysis, to save time, we only gather subreddit data of cryptocurrencies which have a listed subreddit in Coinmarketcap and at least three years of market capitalization of more than 1 million dollars in our sample period. There are 355 cryptocurrenices which pass these criteria. It is worth mentioning that these two extra filters are not considered when constructing the asset pricing factors, and the calculation of idiosyncratic volatility which is derived from the residuals of asset pricing models, is free of look ahead bias. \\
Apart from idiosyncratic volatility and subreddit followers, we also include other cryptocurrency characteristics in our studies in order to analyze the real impact of the variable of interest. In accordance with \cite{ZHANG2020101252} and \cite{yao2021attention}, we calculate and control for the following variables in our empirical model. These characteristics include size (Size): the logarithm of market value; momentum (Mom): the past month’s return; trading volume (Volume): the average logarithm of dollar trading volume for a given cryptocurrency in a specified period; \cite{AMIHUD200231}’s illiquidity measure (equation (\ref{Amihud})): the mean of the absolute daily return divided by the daily dollar trading volume for each month. In equation (\ref{Amihud}), $r_d$ is the daily return rate, $Q_d$ is the daily dollar transaction amount, and $N$ is the number of trading days within a month.
\begin{equation}
\textit{Amihud Illiquidity} = \frac{1}{N}\sum_{d=1}^{N}\frac{|r_d|}{Q_d/10^6}.
\label{Amihud}
\end{equation}
Coinmarketcap classifies cryptocurrencies as coin or token, from which  we try to extract some additional information. Coin refers to any cryptocurrency that has a standalone, independent blockchain, like Bitcoin or Ethereum.  Unlike coins, tokens do not have their own blockchain. Tokens are a unique outlay of broader smart contracts platforms like Ethereum that enable users to create, issue, and manage tokens that are derivatives of the primary blockchain. We define a dummy variable named category which delineates whether a cryptocurrency is a coin or a token.  \\
We use the month to month change in subreddit followers as a proxy to measure the change in investor base. To mitigate possible causality and endogeneity problems, all the independent variables are lagged by one month relative to the dependent variable, as represented by (\ref{regress}). The summary statistics of the cryptocurrencies for which we run the regression are presented in table (\ref{table:summary}).

\begin{table}
    \centering
    \caption{Summary statistics}
\label{tab:02ConTest_GP-PwG}
\begin{tabular*}{\textwidth}{@{\extracolsep{\fill}\quad}@{}cccccccc@{}}
\toprule
& \textbf{}                        & \multicolumn{2}{c}{\multirow{2}{*}{\textbf{\begin{tabular}[c]{@{}c@{}}Number of\\ followers\end{tabular}}}} & \multicolumn{2}{c}{\multirow{2}{*}{\textbf{\begin{tabular}[c]{@{}c@{}}Market capitalization\\ (million dollars)\end{tabular}}}} & \multicolumn{2}{c}{\multirow{2}{*}{\textbf{\begin{tabular}[c]{@{}c@{}}Trading volume\\ (million dollars)\end{tabular}}}} \\
\multicolumn{1}{c}{\multirow{2}{*}{\textbf{\begin{tabular}[c]{@{}c@{}}Year\end{tabular}}}} & \multirow{2}{*}{\textbf{Coins}} & \multicolumn{2}{c}{}                                                                                              & \multicolumn{2}{c}{}                                                                                      & \multicolumn{2}{c}{}                                                                                                                                                                                  \\ \cmidrule(lr){3-8}
\multicolumn{1}{c}{}                                                                                     &                                  & \textbf{Mean}                                         & \textbf{Median}                                         & \textbf{Mean}                                     & \textbf{Median}                                     & \textbf{Mean}                                          & \textbf{Median}                                                                                                                       \\ \midrule
2014&22&11347&840&340&13.08&1.55&0.12\\
 2015&26&12087&1198&173&3.66&1.52&0.04\\
 2016&50&7064&698&215&5.35&2.60&0.07\\
 2017&152&6221&1203&961&68.26&44.50&3.15\\
 2018&275&14240&2502&992&36.50&48.69&1.70\\
 2019&308&15315&2447&658&12.71&114.96&0.96\\
 2020&276&19305&3149&1069&14.76&258.50&1.39\\
 2021&285&41003&4387&5736&70.56&444.37&6.21\\
 2022&246&59386&5940&4916&60.29&287.64&4.07\\
 \bottomrule
\end{tabular*}

\vspace{1ex}
{\justifying \noindent Note: Summary statistics of cryptocurrencies which meet the criteria to be entered into the regression model (equation (6)). Columns related to market capitalization and volume are presented in million dollars.
\par}
\label{table:summary}
\end{table}

\begin{equation}
\textit{IVOL}_{i,t} = \beta_0 + \beta_1*\: (\Delta Investor\: base)_{i,t-1} + \sum_{k}\gamma_k\:Control_{k,i,t-1}+ \sum Cryptocurrency +\sum Period +\epsilon_{i,d}.
\label{regress}
\end{equation}
Here,  $\textit{IVOL}_{i,t}$ is the idiosyncratic volatility of cryptocurrency $i$ in month $t$; $(\Delta Investor\: base)_{i,t-1}$ is the change in investor base of cryptocurrency $i$ in month $t-1$; $Control$s are control variables like size, momentum, trading volume, and Amihud illiquidity  which were defined earlier.  Since we are working with panel data, we must account for possible fixed effects. Fixed effects regression helps us control omitted variables and estimate the effect of intrinsic characteristics of individuals in a panel data set. Omitted variables are not directly observable or measurable but one needs to estimate their effects since forgetting about them leads to a sub-optimally trained regression model. \\
Intuitively, we can think of some possible fixed effects which exist in our panel data. For example, Bitcoin and other high market capitalization cryptocurrencies usually have larger number of followers on social media platforms and consequently, their monthly follower changes would be greater than other cryptos through time. Also, these big cryptocurrencies have usually lower levels of idiosyncratic volatility. Crypto fixed effect is the equivalent of firm fixed effect in stock markets. Based on the nature of data, both idiosyncratic volatility and change in subreddit followers may be subject to serious time or crypto fixed effects. By adding the variables $\sum Cryptocurrency$ and $\sum Period$, we try to account for the existing fixed effects in our panel dataset. \\
In our study, we are mainly interested in the significance of $\beta_1$ and the sign of this coefficient. A significant positive coefficient means that a positive change in investor base increases the cryptocurrency’s idiosyncratic volatility and vice versa. A significant negative coefficient means the opposite.
\section{Empirical Analysis}\label{sec:empirical}

\begin{figure}[htb]
\center{\includegraphics[width=\textwidth]{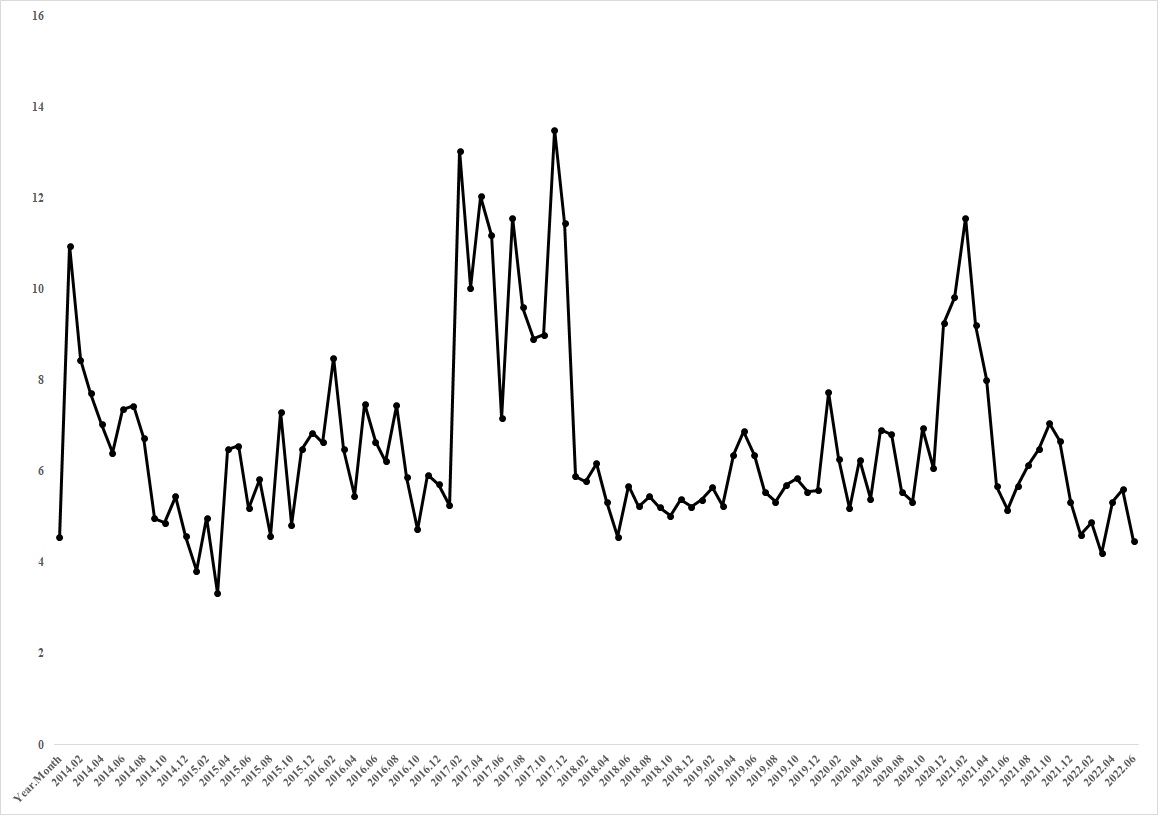}}
\caption{ Time series of the equal-weighted average idiosyncratic volatility in percentage across cryptocurrencies that comprise our sample, from January 2014 to July 2022. For each cryptocurrency, idiosyncratic volatility is estimated as the standard deviation of the residuals in equation (\ref{return}).}
\label{fig:crypto ivol}
\end{figure}

In this section, we present the results. Figure \ref{fig:crypto ivol} plots the equal-weighted average of idiosyncratic volatility across all the cryptocurrencies that comprise the sample for the main regressions of this paper. It is evident from the graph that the average idiosyncratic volatility has been particularly high in 2017 and relatively steady afterwards. 
Table (\ref{table:crypto cr}) reports Pearson correlations between idiosyncratic volatility and other cryptocurrency variables, and table (\ref{table:base change 3factor}) and table (\ref{table:base change CAPM}) present the regression results of equation (\ref{regress}). As could be seen, the coefficient of $(\Delta Investor\: base)_{i,t-1}$ is significantly positive and stays so, even after controlling for a number of control variables. Our results are robust when we switch from the three-factor model to cryptocurrency CAPM. These results suggest that an increase in investor base raises the cryptocurrency’s idiosyncratic volatility. 




\begin{table}
    \centering
    \caption{Correlation matrix of cryptocurrency characteristics}
\label{tab:09RM_Fil_GP-PwG}
\begin{tabular*}{\textwidth}{@{\extracolsep{\fill}\quad}@{}lcccccc@{}}

\toprule
   & \textbf{IVOL 3F} & \textbf{Amihud illiquidity} & \textbf{Momentum} & 
   $\boldsymbol{\Delta}$\textbf{Investor base}& \textbf{Size}& \textbf{Volume}\\ \midrule
\textbf{IVOL 3F}              \textbf{}& 1                                             \\
\textbf{Amihud illiquidity}      & 0.033                & 1                          \\
\textbf{Momentum} & 0.079                & 0.003               & 1             \\
$\boldsymbol{\Delta}$\textbf{Investor base}& -0.029                & 0.002               & 0.061   & 1             \\ 
\textbf{Size}                & 0.212                & -0.008               & 0.034   & 0.320   & 1             \\
\textbf{Volume}        & -0.236                & -0.071               & 0.006     & 0.251 & 0.837 & 1             \\
 \bottomrule
\end{tabular*}


\begin{tablenotes}[flushleft]
\item Note: Pearson correlations between cryptocurrency characteristics. All these variables are defined in the methodology section.
\end{tablenotes}
\label{table:crypto cr}
\end{table}

\begin{table}
    \centering

    \caption{Impact of changes in investor base on cryptocurrencies' idiosyncratic volatility calculated from the three-factor model}
\label{tab:06High30Volat_Fil}
\begin{tabular*}{\textwidth}{@{\extracolsep{\fill}\quad}@{}lcccc@{}}
\toprule
\textbf{}                        & \multicolumn{4}{c}{\textbf{Dependent variable = IVOL 3F}} 
\\ \cmidrule(l){2-5}
\multicolumn{4}{c}{\textbf{\: \:  \: \: \: \: \: \: \: \: \: \: \: \: \: \: \: \: \: \: \: \: \: \: \: \quad \quad \quad \quad \quad \quad \quad \quad \quad \quad \quad \quad Panel A}}
\multirow{2}{*}{}  \\ \cmidrule(l){2-5} 
                                 & \textbf{(1)}                                       & \textbf{(2)}                                       & \textbf{(3)}                                        & \textbf{(4)}                                          \\ \midrule
$\Delta$ Investor\  $\mathrm{base_{t-1}}$    & $0.0000215^{***}$    & $0.0000299^{***}$ & $0.0000294^{***}$     & $0.0000293^{***}$ 

\\
                     & (3.563)                                                 & (5.308)                                                 & (5.226)                                                  & (5.222)                                                 \\
                    $\mathrm{Size_{t-1}}$                              &                                                  & $-0.8874^{***}$                                                 & $-0.5253^{***}$                                                  & $-0.5239^{***}$                                                 \\
                     &                                                & (-15.487)                                                 & (-6.262)                           & (-6.230)                                                 \\ $\mathrm{Volume_{t-1}} $                            &                                                  &                                                  & $-0.3011^{***}$                                                  & $-0.3019^{***}$                                                 \\
                     &                                                  &                                                  & (-5.895)                                                  & (-5.899)                                                 \\ $\mathrm{Amihud\ illiqudity_{t-1}}$                             &                                                  &                                                  &                                                   & -0.3316                                                \\
                     &                                                 &                                                  &                                                   & (-0.257)                                                 \\  Constant                             & $7.5519^{***}$                                                 & $21.4416^{***}$                                                  & $19.2251^{***}$                                                   & $19.2126^{***}$                                                 \\
                     & (63.589)                                                & (22.917)                                                 & (19.103)                                                  & (19.067)                                                 \\
                     Crypto fixed effect                             & Yes                                                 & Yes                                                 & Yes                                                  & Yes                                                 \\  Time fixed effect                             & Yes                                                 & Yes                                                 & Yes                                                  & Yes                                                 \\ Observations                             & 8168                                                 & 7973                                                 & 7973                                                  & 7973                                                 \\ $R^2$ adj                             & 0.283                                                 & 0.302                                                 & 0.306                                                  & 0.305                                                 \\
                      \bottomrule
\end{tabular*}
\begin{tabular*}{\textwidth}{@{\extracolsep{\fill}\quad}@{}lcccc@{}}
\toprule
\textbf{}                        & \multicolumn{4}{c}{\textbf{Panel B}} 
\multirow{2}{*}{}  \\ \cmidrule(l){2-5} 
                                 & \textbf{(5)}                                       & \textbf{(6)}                                       & \textbf{(7)}                                        & \textbf{(8)}                                          \\ \midrule
$\Delta$ Investor\  $\mathrm{base_{t-1}}$     & $0.0000282^{***}$    & $0.0000282^{***}$ & $0.0000286^{***}$     & $0.0000241^{***}$ 

\\
                     & (5.039)                                                 & (5.039)                                                 & (5.110)                                                  & (4.319)                                                 \\
                    $\mathrm{Size_{t-1}}$                              & $-0.6140^{***}$                                                 & $-0.6140^{***}$                                                 & $-0.9443^{***}$                                                                                                  \\
                     & (-7.279)                                                 & (-7.279)                                                 & (-16.457)                                                  &                                                  \\ $\mathrm{Volume_{t-1}}$                             & $-0.2723^{***}$                                                & $-0.2723^{***}$                                                 &                                                   & $-0.5445^{***}$                                                 \\
                     & (-5.334)                                                 & (-5.334)                         &         & (-15.682)                                                 \\ $\mathrm{Amihud\ illiqudity_{t-1}}$                             & $-2.2554^{*}$                                                 & $-2.2554^{*}$                                                 & -1.8796                                                  & $-2.7085^{**}$                                                \\
                     & (-1.729)                                                 & (-1.729)                                                 & (-1.448)                                                  & (-2.073)                                                 \\ $\mathrm{Mom_{t-1}}$                             & $0.4783^{***}$                                                 & $0.4783^{***}$                                                 & $0.4984^{***}$                                                  & $0.4314^{***}$                                                 \\
                     & (8.572)                                                 & (8.572)                                                 & (8.937)                                                  & (7.769)                                                 \\ Category                             &                                                  & -0.0110                                                 &  0.0824                                                 & $-0.3639{**}$                                                 \\
                     &                                                  & (-0.063)                                                 & (0.472)                                                  & (-2.157)                                                 \\ Constant                             & $20.1939^{***}$                                                 & $20.1990^{***}$                                                 & $22.2006^{***}$                                                  & $13.7249^{***}$                                                 \\
                     & (20.007)                                                 & (20.406)                                                 & (24.194)                                                  & (31.329)                                                 \\
                     Crypto fixed effect                             & Yes                                                 & Yes                                                 & Yes                                                  & Yes                                                 \\  Time fixed effect                             & Yes                                                 & Yes                                                 & Yes                                                  & Yes                                                 \\ Observations                             & 7973                                                 & 7973                                                & 7973                                                  & 7973                                                 \\ $R^2$ adj                             & 0.314                                                 & 0.314                & 0.312                                                  & 0.309                                                 \\
                      \bottomrule
\end{tabular*}
\begin{tablenotes}[flushleft]
\item Note: Fixed-effect panel regression estimates for the relation between changes in investor base and cryptocurrencies’ idiosyncratic volatility. All the variables are defined in the methodology section and are monthly. t-statistics are given in the parentheses. *,** and *** denote significance at the 10\%, 5\%, and 1\% levels, respectively. Our dataset covers the period from January 2014 to June 2022.
\end{tablenotes}
\label{table:base change 3factor}
\end{table}

\begin{table}
    \centering

    \caption{Impact of changes in investor base on cryptocurrencies' idiosyncratic volatility calculated from the cryptocurrency CAPM}
\label{tab:06High30Volat_Fil}
\begin{tabular*}{\textwidth}{@{\extracolsep{\fill}\quad}@{}lcccc@{}}
\toprule
\textbf{}                        & \multicolumn{4}{c}{\textbf{Dependent variable = IVOL 1F}} 
\\ \cmidrule(l){2-5}
\multicolumn{4}{c}{\textbf{\: \:  \: \: \: \: \: \: \: \: \: \: \: \: \: \: \: \: \: \: \: \: \: \: \: \quad \quad \quad \quad \quad \quad \quad \quad \quad \quad \quad \quad Panel A}}
\multirow{2}{*}{}  \\ \cmidrule(l){2-5} 
                                 & \textbf{(1)}                                       & \textbf{(2)}                                       & \textbf{(3)}                                        & \textbf{(4)}                                          \\ \midrule
$\Delta$ Investor\  $\mathrm{base_{t-1}}$   & $0.0000241^{***}$    & $0.0000324^{***}$ & $0.0000319^{***}$     & $0.0000318^{***}$ 

\\
                     & (3.674)                                                 & (5.386)                                                 & (5.309)                                                  & (5.304)                                                 \\
                    $\mathrm{Size_{t-1}}$                              &                                                  & $-0.9289^{***}$                                                 & $-0.5696^{***}$                                                  & $-0.5674^{***}$                                                 \\
                     &                                                & (-15.187)                                                 & (-6.358)                                                  & (-6.320)                                                 \\ $\mathrm{Volume_{t-1}}$                             &                                                  &                                                  & $-0.2988^{***}$                                                  & $-0.3001^{***}$                                                 \\
                     &                                                  &                                                  & (-5.480)                                                  & (-5.492)                                                 \\ $\mathrm{Amihud\ illiqudity_{t-1}}$                             &                                                  &                                                  &                                                   & -0.4983                                                \\
                     &                                                 &                                                  &                                                   & (-0.361)                                                 \\  Constant                             & $8.1956^{***}$                                                 & $22.6538^{***}$                                                  & $20.4540^{***}$                                                   & $20.4352^{***}$                                                 \\
                     & (63.520)                                                & (22.684)                                                 & (19.035)                                                  & (18.994)                                                 \\
                     Crypto fixed effect                             & Yes                                                 & Yes                                                 & Yes                                                  & Yes                                                 \\  Time fixed effect                             & Yes                                                 & Yes                                                 & Yes                                                  & Yes                                                 \\ Observations                             & 8168                                                 & 7973                                                 & 7973                                                  & 7973                                                 \\ $R^2$ adj                             & 0.282                                                 & 0.304                                                 & 0.307                                                  & 0.307                                                 \\
                      \bottomrule
\end{tabular*}
\begin{tabular*}{\textwidth}{@{\extracolsep{\fill}\quad}@{}lcccc@{}}
\toprule
\textbf{}                        & \multicolumn{4}{c}{\textbf{Panel B}} 
\multirow{2}{*}{}  \\ \cmidrule(l){2-5} 
                                 & \textbf{(5)}                                       & \textbf{(6)}                                       & \textbf{(7)}                                        & \textbf{(8)}                                          \\ \midrule
$\Delta$ Investor\  $\mathrm{base_{t-1}}$    & $0.0000305^{***}$    & $0.0000305^{***}$ & $0.0000310^{***}$     & $0.0000261^{***}$ 

\\
                     & (5.115)                                                 & (5.115)                                                 & (5.181)                                                  & (4.375)                                                 \\
                    $\mathrm{Size_{t-1}}$                             & $-0.6675^{***}$                                                 & $-0.6675^{***}$                                                 & $-0.9916^{***}$                                                  &                                                  \\
                     & (-7.415)                                                 & (-7.415)                                                 & (-16.197)                                                  &                                                  \\ $\mathrm{Volume_{t-1}}$                             & $-0.2672^{***}$                                                & $-0.2672^{***}$                                                 &                                                   & $-0.5635^{***}$                                                 \\
                     & (-4.904)                                                 & (-4.904)                         &         & (-15.202)                                                 \\ $\mathrm{Amihud\ illiqudity_{t-1}}$                             & $-2.6353^{*}$                                                 & $-2.6353^{*}$                                                 & -2.2686                                                  & $-3.1256^{**}$                                                \\
                     & (-1.893)                                                 & (-1.893)                                                 & (-1.628)                                                  & (-2.24)                                                 \\ $\mathrm{Mom_{t-1}}$                             & $0.5313^{***}$                                                 & $0.5313^{***}$                                                 & $0.5511^{***}$                                                  & $0.4802^{***}$                                                 \\
                     & (8.922)                                                 & (8.922)                                                 & (9.261)                                                  & (8.099)                                                 \\ Category                             &                                                  & 0.0233                                                 & 0.1149                                                  & $-0.3588^{**}$                                                 \\
                     &                                                  & (0.125)                                                 & (0.617)                                                  & (-1.992)                                                 \\ Constant                             & $21.5253^{***}$                                                 & $21.5144^{***}$                                                 & $23.4785^{***}$                                                  & $14.4818^{***}$                                                 \\
                     & (19.981)                                                 & (20.364)                                                 & (23.980)                                                  & (30.959)                                                 \\
                     Crypto fixed effect                             & Yes                                                 & Yes                                                 & Yes                                                  & Yes                                                 \\  Time fixed effect                             & Yes                                                 & Yes                                                 & Yes                                                  & Yes                                                 \\ Observations                             & 7973                                                 & 7973                                                & 7973                                                  & 7973                                                 \\ $R^2$ adj                             & 0.314                                                 & 0.314                & 0.312                                                  & 0.309                                                 \\
                      \bottomrule
\end{tabular*}
\begin{tablenotes}[flushleft]
\item Note: Fixed-effect panel regression estimates for the relation between changes in investor base and cryptocurrencies’ idiosyncratic volatility. All the variables are defined in the methodology section and are monthly. t-statistics are given in the parentheses. *,** and *** denote significance at the 10\%, 5\%, and 1\% levels, respectively. Our dataset covers the period from January 2014 to June 2022.
\end{tablenotes}
\label{table:base change CAPM}
\end{table}

Also, the coefficients of control variables are in line with  previous studies (\cite{yao2021attention}). For instance, the coefficient of size and volume are always significantly negative. This means that larger scale cryptocurrencies have lower levels of idiosyncratic volatility. Moreover, the coefficient of momentum is significantly positive, which states that higher past months returns lead to higher levels of idiosyncratic volatility. It is also worth noting that by adding size or volume, the coefficient of $(\Delta Investor\: base)_{i,t-1}$ becomes larger. It may be that adding these two variables absorbs some of the residual noise and results in a more precise estimation of the coefficient of our variable of interest. \\
The dummy variable added for category is insignificant in presence of size. This result is not suprising, since coins generally have higher market capitalization than tokens. This variable is $1$ if the cryptocurrency is a coin and $0$ if it is a token, it is only considered to see whether this characteristic has any effect on the idiosyncratic volatility. \\ 
From the correlation matrix in table (\ref{table:summary}), it is clear that size and volume are highly correlated. This may raise concerns about the presence of multicollinearity in our regressions. Multicollinearity can be detected using various techniques, one such technique being the Variance Inflation Factor (VIF). In the VIF method, we pick each feature and regress it against all of the other features. For each regression, the factor is calculated as:
\begin{equation}
VIF = \frac{1}{1-R^2}.
\label{VIF}
\end{equation}
Table (\ref{table:collinearity}) shows the VIF values for the main variables of columns (6), (7) and (8) of tables (\ref{table:base change 3factor}) and (\ref{table:base change CAPM}). The VIF values for size and volume when simultaneously present as control variables, are above the critical value of 10 and indicate the presence of high multicollinearity. But when only size or volume is present, the VIF values drop below the critical value. Moreover our tabulated results do not display the typical symptoms of multicollinearity such as drastic changes of coefficients when other control variables are added. Thus we could conclude that the results our not subject to multicollinearity.

\begin{table}
    \centering

    \caption{Test of multicollinearity}
\label{tab:06High30Volat_Fil}
\begin{tabular*}{\textwidth}{@{\extracolsep{\fill}\quad}@{}lccc@{}}
\toprule
\textbf{}                        & \multicolumn{3}{c}{\textbf{Variance inflation factor}} 
\\ \cmidrule(l){2-4}

                                 & \textbf{(1)}                                       & \textbf{(2)}                                       & \textbf{(3)}                                                       \\ \midrule
$\Delta$ Investor  base    & 1.36   & 1.35 & 1.37     

        \\
                     Size                            &            18.22                                      &                                            & 8.41                                                                                \\ Volume                             &  14.03                                                &   6.48                                               &                                                                                           \\ Amihud illiqudity                             &     1.09                                             &                                    1.09              &                 1.09                                                                    \\  Momentum                             & 1.38                                                 & 1.36                                                  & 1.37                                 \\
                     
                      \bottomrule
\end{tabular*}
\begin{tablenotes}[flushleft]
\item Note: This table reports the VIF values of main variables . VIF values above 10 indicate the presence of high multicolinearity. columns (1), (2) and (3) of this table correspond to columns (6), (7) and (8) of tables (4) and (5), respectively. 
\end{tablenotes}
\label{table:collinearity}
\end{table}

\newpage
\section{Conclusion}\label{sec:conclusion}

Although cryptocurrencies are characterized by substantial volatility and many studies (\cite{merton1987model}, \cite{malkielxu2004}) have pointed out that in the case of market friction investors require compensation for securities’ idiosyncratic volatility, little research has been conducted on idiosyncratic volatility of cryptocurrency markets. In addition, recent incidents like GameStop short squeeze coordinated by Reddit users exhibit the growing influence of social media on the dynamics of financial markets. This phenomenon seems to be only worse in cryptocurrency markets. This paper tries to link these two strands of research.  \\
We use cryptocurrency CAPM and Fama-French three-factor model to calculate different measures of cryptocurrency idiosyncratic volatility. Then, by utilizing change in subreddit followers of each cryptocurrency as a proxy for the change in investor base, we find that changes in investor base can significantly increase cryptocurrencies’ idiosyncratic volatility. Our findings remain significant even after controlling for a number of cryptocurrency characteristics, namely size, momentum, liquidity and volume. Our conclusion contributes to the growing literature of idiosyncratic volatility of cryptocurrencies and provides another evidence as to how social media platforms, especially Reddit, can affect the functioning of financial markets.

\clearpage

\setcitestyle{numbers} 
\bibliographystyle{plainnat}
\bibliography{refs}

\clearpage

\end{document}